# SynConfRoute: Syntax-Aware Routing for Efficient Code Completion with Small CodeLLMs

Kishanthan Thangarajah
Centre for Software Excellence,
Huawei Canada
Canada
cse@huawei.com

Boyuan Chen
Centre for Software Excellence,
Huawei Canada
Canada
cse@huawei.com

Ahmed E. Hassan
Queen's University
Kingston, Canada
ahmed@cs.queensu.ca

## Abstract

Enterprises want AI code completion that is both high-quality and private, but they face a tension: proprietary models yield better results yet risks exposing proprietary code, while self-hosting large models is expensive and hard to maintain. As a lighter alternative, small CodeLLMs (1 B–3 B) can run on a developer's workstation accelerator with code never leaving the machine, but they fail on harder tasks. A practical solution is to use the small model for most requests and selectively route difficult ones to a larger self-hosted model. In this study, we evaluate 29 code specialized LLMs (0.5 B–480 B) from 12 families on execution-based fill-in-the-middle (FIM) code completion benchmarks across Python, Java, and C++, and find that model family and code specialized training matter more than size: a 3 B model matches a 32 B model despite being 10× smaller. Analyzing the 3 B model's failures, we discover that 46% of its incorrect completions are not valid code. To enable efficient code completion, we propose *SynConfRoute*, a training-free method that combines token confidence with syntax validation to automatically decide per-request whether to keep the local completion or escalate to a larger self-hosted larger model. *SynConfRoute* improves pass@1 by 6.4% over confidence only routing on routine completions and by up to 31% on harder multi-language tasks, and the resulting pipeline achieves 78.9% on routine completions, 7.4% higher than always using the 480 B model alone, while reducing accelerator usage by 58%. *SynConfRoute* generalizes across Python, Java, and C++, improving over confidence only routing on all three languages without ever rejecting a correct local completion. The pipeline uses off-the-shelf models with no custom training, making it immediately deployable in practice.

## 1 Introduction

AI-generated code is becoming the norm: Google reports that 75% of its new code is now AI-generated [38] and Anthropic reports 70–90% [7]. Much of this is driven by inline code completion, the most widely adopted form of AI coding assistance, with 84% of developers using or planning to use such tools [45]. However, organizations face a tension between quality and data privacy. The most capable CodeLLMs (30B to 480B+) require accelerator infrastructure that is either expensive to self-host or accessed via external APIs that risk exposing proprietary source code. This risk is not hypothetical: Samsung banned all generative AI tools after engineers leaked confidential semiconductor code through ChatGPT [36], and a Cisco survey found that 27% of organizations have banned generative AI use over privacy and data security concerns [8]. Regulatory pressures such as GDPR [14] add further urgency. Industry is already responding: Meta deploys a fine-tuned CodeLlama for internal code completion [31], JetBrains ships lightweight on-device models for low-latency line-level completions [43], and organizations increasingly self-host CodeLLMs to address both privacy and latency concerns [10, 32, 48].

Small CodeLLMs (1B to 3B parameters) offer an appealing local-first alternative: they can run on a developer's workstation with a small accelerator at sub-second latency, keeping code fully on-device with no data exposure risk. The question is whether they are good enough for most completions, and for the tasks where they fall short, whether selectively routing to a larger model on the organization's shared accelerator can fill the gap without routing every request through expensive infrastructure.

**Motivating example.** To illustrate the challenge, we ran six CodeLLMs of increasing size (primarily from the Qwen family to control for training differences) on six FIM tasks of increasing difficulty (Table 1). The results reveal a clear pattern: sub-1B models almost always fail, Qwen2.5-Coder-3B (hereafter *3B*) handles three of six tasks correctly, and only the 7B model and the Qwen3-Coder-480B-A35B (*480B*, a Mixture-of-Experts model with 35B active parameters) solve all six. The 3B model's failures involve subtle errors such as using the wrong variable name, producing incomplete logic, and omitting a recursive call. These are knowledge gaps, not formatting issues. The practical challenge is to identify *which* tasks the local model can handle, per-request, at inference time.

**Table 1: Motivating example: fill-in-the-middle completions across model sizes. All Qwen2.5-Coder except SC-1B (StarCoder-1B) and 480B (Qwen3-Coder-480B MoE). Diff. = Difficulty. ✓ = correct, × = incorrect.**

| Task | Diff. | 0.5B | SC-1B | 1.5B | 3B | 7B | 480B |
|---|---|---|---|---|---|---|---|
| Variable assign. | Easy | ✓ | × | ✓ | ✓ | ✓ | ✓ |
| List filtering | Easy | × | × | ✓ | × | ✓ | ✓ |
| Dict update | Med. | × | × | × | ✓ | ✓ | ✓ |
| Binary search | Med. | × | × | × | ✓ | ✓ | ✓ |
| Recursive flatten | Hard | × | × | ✓ | × | ✓ | ✓ |
| Memoize decorator | Hard | × | × | × | × | ✓ | ✓ |
| **Total** | | 1/6 | 0/6 | 3/6 | 3/6 | 6/6 | 6/6 |

This question has become increasingly tractable as code specialized models have matured. For example, Qwen2.5-Coder [21], trained on over 5.5 T tokens with FIM objectives, has dramatically raised the quality ceiling for sub-10 B models. Yet it remains unclear which models offer the best trade-off between quality and cost, when local serving is sufficient versus when escalation is needed, and whether inference-time techniques can close the remaining quality gap. While model routing has been studied for general natural language processing (NLP) tasks [5, 35, 54], code completion differs fundamentally: correctness is determined by whether the generated code passes execution-based tests, not by preference or surface similarity, and routing strategies effective for NLP may not transfer. We address these through three research questions:

- **RQ1: Which small CodeLLM should organizations deploy locally?** We benchmark 29 CodeLLMs (0.5 B to 480 B) from 12 model families on execution-based FIM tasks across quality, latency, and quantization.
- **RQ2: Can code-specific signals improve routing beyond confidence?** We analyze failure modes of confident-but-wrong completions and propose a routing method that exploits structural properties unique to code.
- **RQ3: Where does local-first routing work and where does it break down?** We test *SynConfRoute* on harder benchmarks across multiple languages, evaluate whether inference-time alternatives can substitute for routing, and identify the remaining quality gap.

To our knowledge, this is the first systematic study of model routing for code completion using execution-based evaluation. We evaluate on two FIM benchmarks (HumanEval-Infilling [3] and SAFIM [17]), compare six baseline routing methods, and propose *SynConfRoute* as a code-specific alternative.

Our principal findings are:

- **Comprehensive model selection (RQ1).** We evaluate 29 code specialized LLMs from 12 families. Model family and training objective matter more than size: Qwen2.5-Coder-3B achieves 64.7% pass@1 at 0-shot, matching the 32B model despite being 10× smaller, while an 80B agentic model (Qwen3-Coder-Next) scores only 41.1% because it was not trained for FIM completion. Q4_K_M quantization retains full fp16 quality; sub-1B models and aggressive quantization prove non-viable.
- **Code-specific failure decomposition enables better routing (RQ2).** We discover that confident-but-wrong completions decompose into structurally broken (46%, detectable via syntax checking) and semantically wrong (54%, requiring model improvements). *SynConfRoute* exploits this, improving pass@1 by 6.4% over confidence only routing on routine completions (78.9% vs 72.5%) and by up to 31% on harder multi-language tasks, with zero false positives. *SynConfRoute* automatically keeps the 3B model's correct completions local (including cases where it outperforms the 480B model) and escalates only genuinely difficult tasks, keeping 58% of requests on-device.
- **Generalizes across languages with clear boundaries (RQ3).** The pipeline achieves +7.4% quality over the large model alone while reducing accelerator usage by 58% on routine tasks. *SynConfRoute* generalizes across Python, Java, and C++, improving on every SAFIM benchmark language and task type without ever rejecting a correct completion. The boundary is error type (structural vs. semantic), not language. The pipeline uses off-the-shelf models with no custom training, making it immediately deployable.

The remainder of the paper is organized as follows. Section 2 describes benchmarks, models, and evaluation setup. Sections 3 to 5 present results. Section 6 discusses practical implications. Section 7 reviews related work. Section 8 addresses threats to validity, and Section 9 concludes.

## 2 Methodology

This section describes the benchmarks, models, prompting strategy, routing methods, and inference-time strategies used in our study.

### 2.1 Benchmarks

We use FIM as the evaluation paradigm, which directly models IDE code completion: given code before the cursor (prefix) and after the cursor (suffix), the model predicts the missing fragment. We evaluate on two execution-based FIM benchmarks.

**HumanEval-Infilling** [3] derives FIM tasks from the 164 HumanEval functions by masking individual lines. We use the single-line split (1,033 tasks). Evaluation uses pass@1: the completed program is executed against unit tests, and a task is solved if all tests pass.

**SAFIM** [17] provides syntax-aware FIM tasks from competitive programming submissions. Tasks are categorized into three subtypes: *control-flow* completion, *algorithmic block* completion, and *API function call* completion. SAFIM covers four languages; we evaluate on Python (750 control, 805 block, 181 API), Java (2,464 control, 2,479 block, 56 API), and C++ (4,881 control, 4,968 block, 52 API). RQ1 uses all three languages for model selection (Table 3); RQ2 uses Python (HumanEval-Infilling) for routing method development; RQ3 validates across all three languages. Evaluation uses pass@1 via the ExecEval [23] framework.

We chose these two benchmarks as the only publicly available execution-based FIM benchmarks with deterministic test suites, ensuring correctness is measured by execution rather than surface similarity. We additionally report inference latency (median wall-clock time per completion) on accelerator hardware.

**Table 2: Models evaluated. Small and mid-range models are served via Ollama; large models via vLLM.**

| Model | Family | Size | Serving |
|---|---|---|---|
| *Small (≤3 B)* | | | |
| Qwen2.5-Coder-0.5B | Qwen | 0.5B | Ollama |
| StarCoder-1B | StarCoder | 1B | Ollama |
| DeepSeek-Coder-1.3B | DeepSeek | 1.3B | Ollama |
| Qwen2.5-Coder-1.5B | Qwen | 1.5B | Ollama |
| Yi-Coder-1.5B | Yi-Coder | 1.5B | Ollama |
| OpenCoder-1.5B | OpenCoder | 1.5B | Ollama |
| Refact-1.6B | Refact | 1.6B | Ollama |
| CodeGemma-2B | CodeGemma | 2B | Ollama |
| Qwen2.5-Coder-3B | Qwen | 3B | Ollama |
| StarCoder2-3B | StarCoder | 3B | Ollama |
| Stable-Code-3B | Stable | 3B | Ollama |
| Granite-Code-3B | Granite | 3B | Ollama |
| *Mid-range (6.7 B to 16 B)* | | | |
| DeepSeek-Coder-6.7B | DeepSeek | 6.7B | Ollama |
| Qwen2.5-Coder-7B | Qwen | 7B | Ollama |
| StarCoder2-7B | StarCoder | 7B | Ollama |
| CodeLlama-7B | CodeLlama | 7B | Ollama |
| CodeGemma-7B | CodeGemma | 7B | Ollama |
| Magicoder-7B | Magicoder | 7B | Ollama |
| OpenCoder-8B | OpenCoder | 8B | Ollama |
| Granite-Code-8B | Granite | 8B | Ollama |
| Yi-Coder-9B | Yi-Coder | 9B | Ollama |
| CodeGeeX4-9B | CodeGeeX | 9B | Ollama |
| DS-Coder-V2-Lite | DeepSeek | 16B MoE | Ollama |
| *Large reference (accelerator-served)* | | | |
| Qwen3-Coder-30B-A3B | Qwen | 30B MoE | vLLM |
| Qwen2.5-Coder-32B | Qwen | 32B | vLLM |
| DeepSeek-Coder-33B | DeepSeek | 33B | vLLM |
| CodeLlama-34B | CodeLlama | 34B | vLLM |
| Qwen3-Coder-Next | Qwen | 80B MoE | vLLM |
| Qwen3-Coder-480B-A35B | Qwen | 480B MoE | vLLM |

## 2.2 Models

We evaluate 29 CodeLLMs spanning 12 model families (23 small and mid-range, 6 large reference) as listed in Table 2. We selected all publicly available code specialized LLMs with Ollama-compatible GGUF weights at sizes from 0.5B to 16B, plus six large models as references. We focus on code specialized models rather than general-purpose LLMs (e.g., Llama, Mistral) because FIM completion benefits from dedicated fill-in-the-middle training objectives [3] that general-purpose models typically lack. Our own results confirm this: even Qwen3-Coder-Next, a code model trained for agentic tasks rather than FIM, achieves only 41.1% despite having 80B parameters (Section 3). Small and mid-range models are served via Ollama [34] with Q4_K_M quantization on a single 80 GB accelerator. The recommended 3 B model at Q4_K_M requires only ∼2 GB of accelerator memory, fitting any workstation accelerator with 8 GB or more; we use a single 80 GB accelerator for benchmarking all 23 models uniformly. All six large reference models are served via vLLM [24] on the same configuration of 8×80 GB accelerators with tensor parallelism, ensuring fair latency comparisons across models. We additionally conduct a quantization sensitivity study across four levels (q2_K, Q4_K_M, q8_0, fp16) for the 1.5B and 3B models, spanning the full range from aggressive 2-bit to unquantized, with Q4_K_M and q8_0 as the two most commonly deployed intermediate levels. We selected 1.5B and 3B as the best-performing models at their respective size tiers and the primary local deployment candidates.

## 2.3 Prompting Strategy

All models use a unified chat-based infilling prompt that provides the code prefix and suffix as context and instructs the model to return only the missing code, rather than native FIM tokens, enabling fair comparison across all 29 models. In preliminary experiments, Qwen2.5-Coder-3B via chat prompting (64.7%) outperformed all native FIM configurations, including the same model's own FIM mode (55.2%) and other families (26% to 35%), confirming that instruction-following quality matters more than FIM token support for single-line completion. A post-processing indentation repair step is applied uniformly to all models, as chat-based completions often produce incorrect indentation levels that cause test failures despite correct logic. All experiments use greedy decoding (temperature 0.0) for deterministic reproducibility, as is standard for pass@1 evaluation [6]. We set a 50-token generation limit because single-line completions in our benchmarks average 8–12 tokens, providing a 4–6x margin.

## 2.4 Routing Methods

We evaluate six baseline routing methods spanning two categories.

**Pre-inference methods** decide before the local model runs: (i) *Static code signals*, a decision tree (max depth 4) on four features (prefix/suffix length, nesting depth, identifier overlap); (ii) *Embedding KNN* (K=11) using all-MiniLM-L6-v2 [41]; (iii) *Combined*, static features concatenated with embeddings; and (iv) *Eagle ELO routing* (binary and ternary variants) [54], a training-free method using ELO ratings from calibration data.

**Post-inference methods** decide after the local model runs: (v) *Confidence-only*, escalates if the average log-probability of the first three generated tokens falls below a threshold, as proposed by MCCom [28]; and (vi) *Cascade*, which combines a low-confidence gate with a second threshold for borderline cases.

All methods are calibrated on a 200-task subset (seed 42) from HumanEval-Infilling and evaluated on the remaining 833 test tasks.

## 2.5 Inference-Time Strategies

Beyond routing, we evaluate two inference-time strategies for improving small model quality without retraining: (i) *Few-shot retrieval*, five strategies for selecting a 1-shot example (0-shot baseline, static, random, TF-IDF (term frequency-inverse document frequency) retrieval, and embedding retrieval); and (ii) *Multi-sample voting*, generating $N \in \{3, 5, 7, 10\}$ completions and selecting the majority output. Full experimental details are provided in Section 5.

# 3 RQ1: Which Small CodeLLM Should Organizations Deploy Locally?

The motivating example (Table 1) suggests that a well-trained small model can match larger models on routine tasks. We expect this pattern to be systematic, with model family and code specialized training mattering more than raw parameter count, and moderate quantization retaining most quality for local deployment.

## 3.1 Experimental Setup

We evaluate all 29 CodeLLMs (Table 2) on HumanEval-Infilling (1,033 single-line Python tasks) and SAFIM (1,736 Python tasks

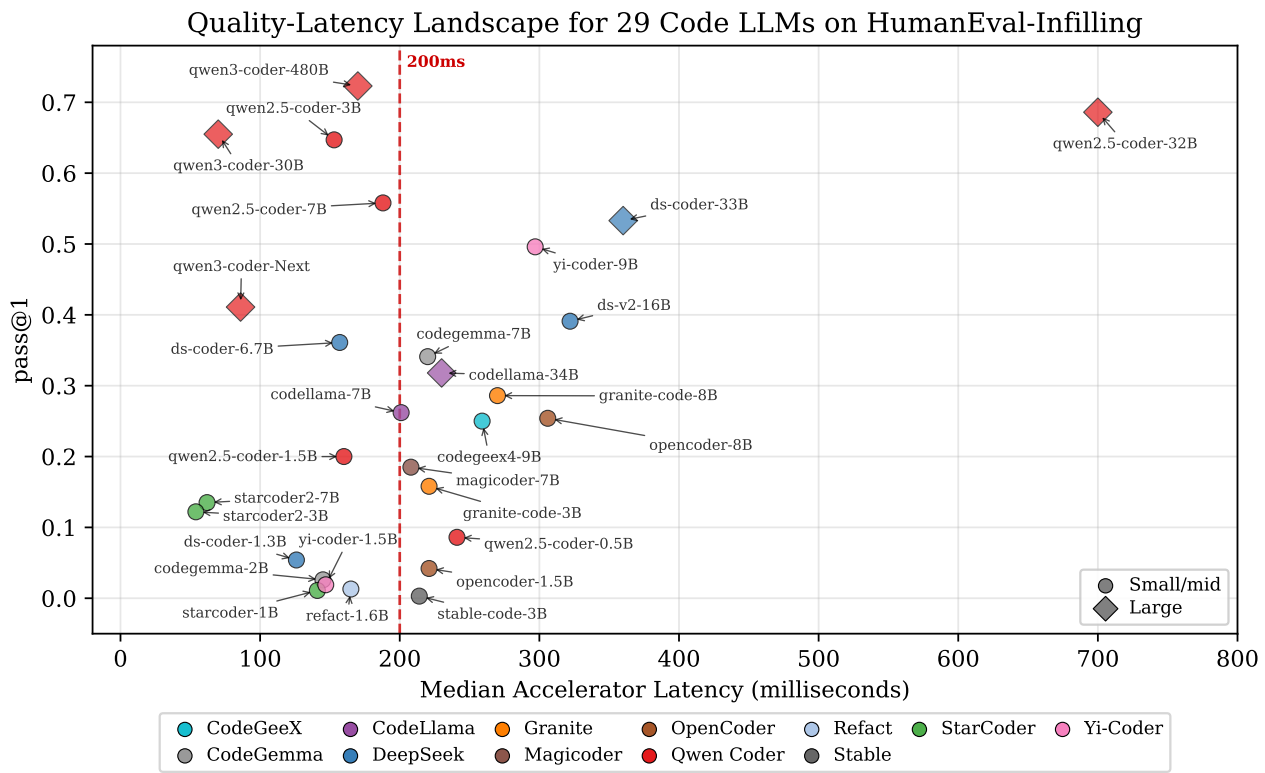

**Figure 1: Quality-latency landscape for 29 CodeLLMs on HumanEval-Infilling single-line. All models on accelerator. Circles = small/mid models, diamonds = large models. The dashed line marks the 200 ms real-time target.**

across three subtypes, plus Java and C++ for multi-language generalization). Each model is tested at 0-shot and 1-shot; five of the six large models are additionally swept from 0-shot to 5-shot (Qwen3-Coder-Next is evaluated at 0-shot only). Small and mid-range models run via Ollama at Q4_K_M quantization on a single 80 GB accelerator; large models run on accelerator via vLLM. We report pass@1 (execution-based correctness) and median inference latency per completion.

## 3.2 HumanEval-Infilling Results

Figure 1 plots pass@1 (best configuration) against median accelerator latency for all 29 models, with families distinguished by color and large reference models marked as diamonds.

**Model family dominates size.** Qwen2.5-Coder-3B achieves 64.7% pass@1 at 0-shot, within 0.1% of Qwen2.5-Coder-32B (64.8%) and ahead of Qwen3-Coder-30B (64.2%). By contrast, models of comparable or larger size from other families perform significantly worse: CodeLlama-7B (26.2%), OpenCoder-8B (25.4%), Granite-Code-8B (28.6%), CodeGeeX4-9B (25.0%), and even CodeLlama-34B (31.8%), a model 11× larger, likely due to weaker FIM training data and older training recipes. This dominance holds on SAFIM: the 3B achieves 23.1% on block tasks versus 13.5% for CodeGeeX4-9B, 5.5% for CodeGemma-7B, and 1.2% for Yi-Coder-9B.

**Training objective matters more than scale.** Qwen3-Coder-Next, an 80B MoE model with 3B active parameters trained for agentic coding, achieves only 41.1% on HumanEval-Infilling despite its larger capacity. The FIM-trained Qwen2.5-Coder-3B outperforms it by 23.6%, confirming that a model trained for the target task (FIM completion) outperforms a larger model trained for a different objective (agentic coding).

**Size inversion within the same family.** Within Qwen2.5-Coder, the 3B model (64.7%) outperforms the 7B model (54.0%) by 10.7%. This result is consistent across 0-shot and 1-shot configurations and persists at fp16 quantization (64.4% vs 54.0%), ruling out quantization as the cause.

**Table 3: SAFIM pass@1 across Python, Java, and C++ for four models. Unlike HumanEval-Infilling where the 3B and 32B appear equivalent, SAFIM reveals a clear quality gap that motivates routing.**

| | Python | | | Java | | | C++ | | |
|---|---|---|---|---|---|---|---|---|---|
| Model | Ctrl | Blk | API | Ctrl | Blk | API | Ctrl | Blk | API |
| 1.5B | .000 | .009 | .028 | .008 | .024 | .089 | .007 | .005 | .000 |
| 3B | .007 | .231 | .260 | .032 | .189 | .232 | .024 | .098 | .058 |
| 32B | .153 | .327 | .586 | .272 | .180 | .714 | .234 | .106 | .519 |
| 480B | .461 | .417 | .674 | .473 | .324 | .893 | .423 | .221 | .731 |

**1-shot helps most models but hurts the best.** Most small and mid-range models improve with 1-shot prompting, with the largest gains for Yi-Coder-9B (+24.9%), Granite-Code-8B (+28.3%), and OpenCoder-8B (+14.6%). However, Qwen2.5-Coder-3B drops from 64.7% to 56.8% (−7.9%). Failure case analysis shows the 3B model copies the example's indentation and code patterns rather than solving the actual task, an effect investigated further in Section 5.

**Half of the small models meet 200 ms on accelerator.** On a single 80 GB workstation accelerator, the 3B model completes in 0.15 s (Figure 1). Of the 23 small/mid models evaluated, 11 meet the 200 ms real-time target, including the recommended 3B model. The quality gaps revealed by SAFIM below, not latency, motivate the routing study in RQ2.

**Few-shot prompting helps weak models but hurts the best.** Among large models, Qwen3-Coder-480B peaks at 0-shot (72.3%) and degrades with examples (−4.0% at 1-shot), matching the 3B pattern. Qwen2.5-Coder-32B improves steadily to 4-shot (+3.8%), and DeepSeek-Coder-33B benefits most (+14.3% to 53.3%).

## 3.3 SAFIM Results

Table 3 presents SAFIM results across Python, Java, and C++ for the four models used in subsequent routing and retrieval experiments.

**SAFIM is more discriminating than HumanEval-Infilling.** The 3B and 32B models appear equivalent on HumanEval-Infilling (64.7% vs 64.8%), but on SAFIM block tasks a clear gap emerges: 23.1% vs 32.7% (Δ9.6%). The 480B model extends this advantage further, reaching 46.1% on control, 41.7% on block, and 67.4% on API.

**Task type strongly affects difficulty.** API completion is the easiest subtask across all three languages. Control-flow is hardest in Python (32B: 15.3%), while block is hardest in Java and C++.

**The 1.5B model is near-zero across all languages.** Python scores of 0% (control), 0.9% (block), and 2.8% (API) are mirrored in Java and C++, confirming that the quality gap motivating routing is not language-specific.

**C++ is systematically harder than Java.** The 480B model scores 32.4% on Java block vs 22.1% on C++ block.

## 3.4 Quantization Sensitivity

We evaluate four quantization levels (q2_K, Q4_K_M, q8_0, fp16) on the 1.5B and 3B models. Q4_K_M retains full fp16 quality for the 3B model (64.7% vs 64.4%) at the lowest latency, while Q2_K is catastrophic (0.0% for 3B). Practitioners should use Q4_K_M for models ≥3B and q8_0 for smaller models. Sub-1B models (StarCoder-1B: 1.1%) are non-viable regardless of quantization.

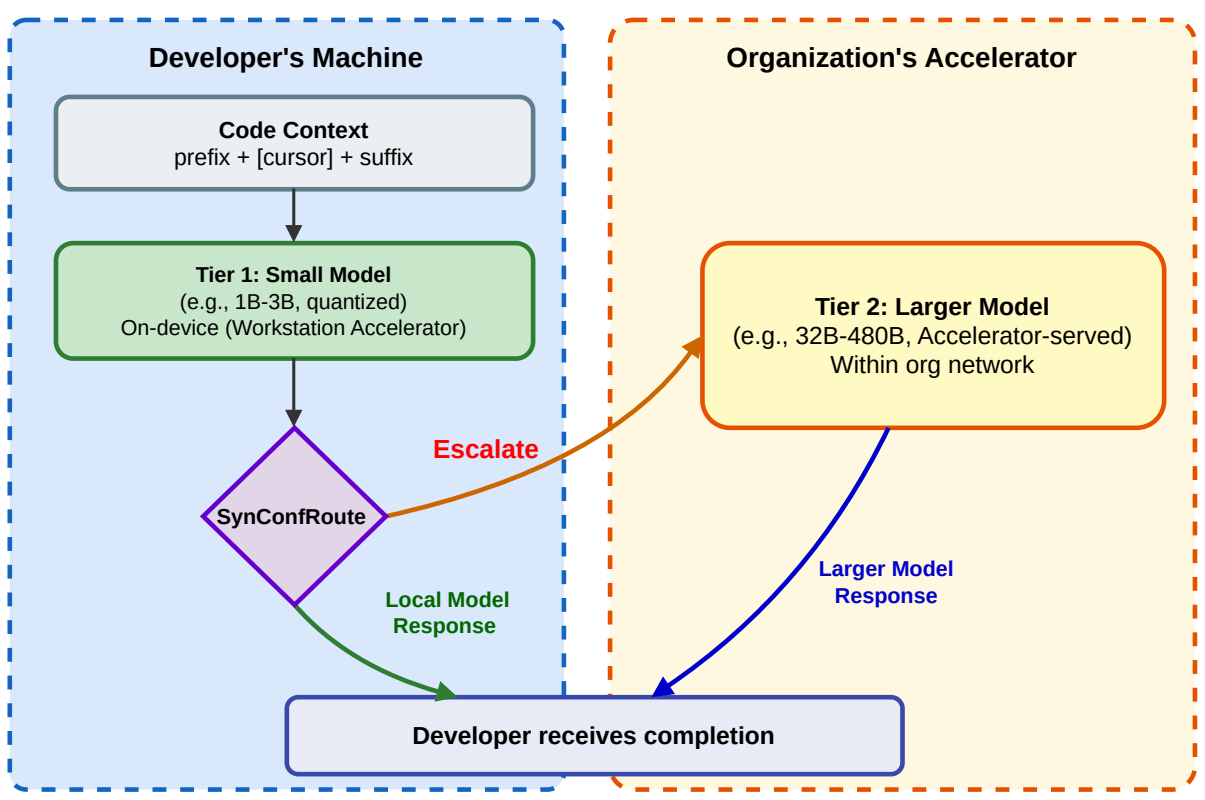

**Figure 2: Local-first deployment architecture. The small model runs on the developer's workstation accelerator (code stays on-device).** ***SynConfRoute*** **determines per-request whether to keep the completion local or escalate to a larger model on the organization's shared accelerator.**

> **RQ1 Conclusion:** A well-chosen 3B model handles routine completions as well as 30B+ models (64.7% pass@1), and Q4_K_M retains full quality for on-device deployment. However, SAFIM reveals a 9.6% quality gap on harder tasks that HumanEval-Infilling masks entirely.
>
> **Implication:** Practitioners should deploy by model family, not size. The remaining quality gap on harder tasks motivates code-specific routing (RQ2).

## 4 RQ2: Can Code-Specific Signals Improve Routing Beyond Confidence?

Given the quality gap on harder tasks (RQ1), we investigate whether code-specific validation signals can improve routing beyond confidence only methods used in NLP. Code has a property NLP text lacks: structural correctness can be cheaply verified.

### 4.1 Local-First Deployment Architecture

Figure 2 illustrates the local-first deployment architecture. A small model (1.5B or 3B) runs on the developer's workstation accelerator, keeping code on-device. A larger model (32B or 480B) is available on the organization's shared accelerator infrastructure for escalation. The router decides per-request whether to accept the local completion or escalate. This two-tier design balances three objectives: quality (by leveraging the larger model for hard tasks), privacy (by keeping routine completions local), and cost (by reducing shared accelerator usage).

**Algorithm 1** SynConfRoute

**Require:** Task $x$ with prefix $p$, suffix $s$; local model $M_L$; larger model $M_R$; threshold $t^*$

1: $\hat{y}, c_x \leftarrow M_L(x)$ {generate completion + confidence}
2: **if** $c_x < t^*$ **then**
3: **return** $M_R(x)$ {low confidence → escalate}
4: **end if**
5: **if** $\neg\,\mathrm{SyntaxValid}(p \oplus \hat{y} \oplus s)$ **then**
6: **return** $M_R(x)$ {syntax invalid → escalate}
7: **end if**
8: **return** $\hat{y}$ {confident + valid → keep local}

**Threshold:** $t^* = \arg\max_t \text{pass@1}(t)$, selected by grid search on a held-out set of ≥50 tasks.

### 4.2 SynConfRoute: Confidence + Syntax Routing

We propose *SynConfRoute*, which extends confidence only routing with a code-specific signal: syntax validation. The key observation is that code, unlike natural language, has a cheaply verifiable structural property: syntactic correctness. If a completion does not parse, it cannot be functionally correct. This gives us a zero-cost filter that is unavailable in NLP routing.

*SynConfRoute* works in two steps (Algorithm 1). First, the local model generates a completion and a confidence score, measured as the average log-probability of the first three generated tokens [28]. If confidence falls below a threshold $t^*$, the request is escalated to the larger model. Second, if the model is confident, the completed code (prefix + completion + suffix) is checked for syntactic validity using a language-specific parser: Python's `ast.parse` (<1 ms), Java's `javalang` (<1 ms), or `g++ -fsyntax-only` for C++ (~150 ms). If the completion is syntactically invalid, the request is escalated regardless of confidence. Only completions that are both confident and syntactically valid are kept local. The threshold $t^*$ is selected by grid search on a small held-out set of ≥50 tasks.

### 4.3 Experimental Setup

We evaluate routing with two local models (1.5B and 3B) paired with two larger models (32B and 480B), all from the Qwen family to isolate routing effectiveness from tokenizer mismatches. The six baseline methods (Section 2.4) and *SynConfRoute* (Algorithm 1) are calibrated on a 200-task subset and evaluated on the remaining 833 test tasks. Because of this split, pass@1 numbers in RQ2 differ slightly from RQ1 (which reports on all 1,033 tasks): the 3B model scores 63.3% on the 833-task test set vs 64.7% overall, and the 480B scores 71.5% vs 72.3%. Results with 32B are presented first; Section 4.4 evaluates the 480B MoE model.

### 4.4 Routing Results

Figure 3 compares all routing methods on the recommended 3B+480B pair. We evaluate with both the 32B dense model and the 480B MoE as escalation targets; the 480B is the recommended target because it achieves higher quality (72.5% vs 70.0% for confidence routing) while being faster (0.17 s vs 0.70 s on the same 8-accelerator setup) due to sparse activation.

**Confidence routing surpasses both individual models.** As shown in Figure 3, confidence routing at threshold 0.7 achieves

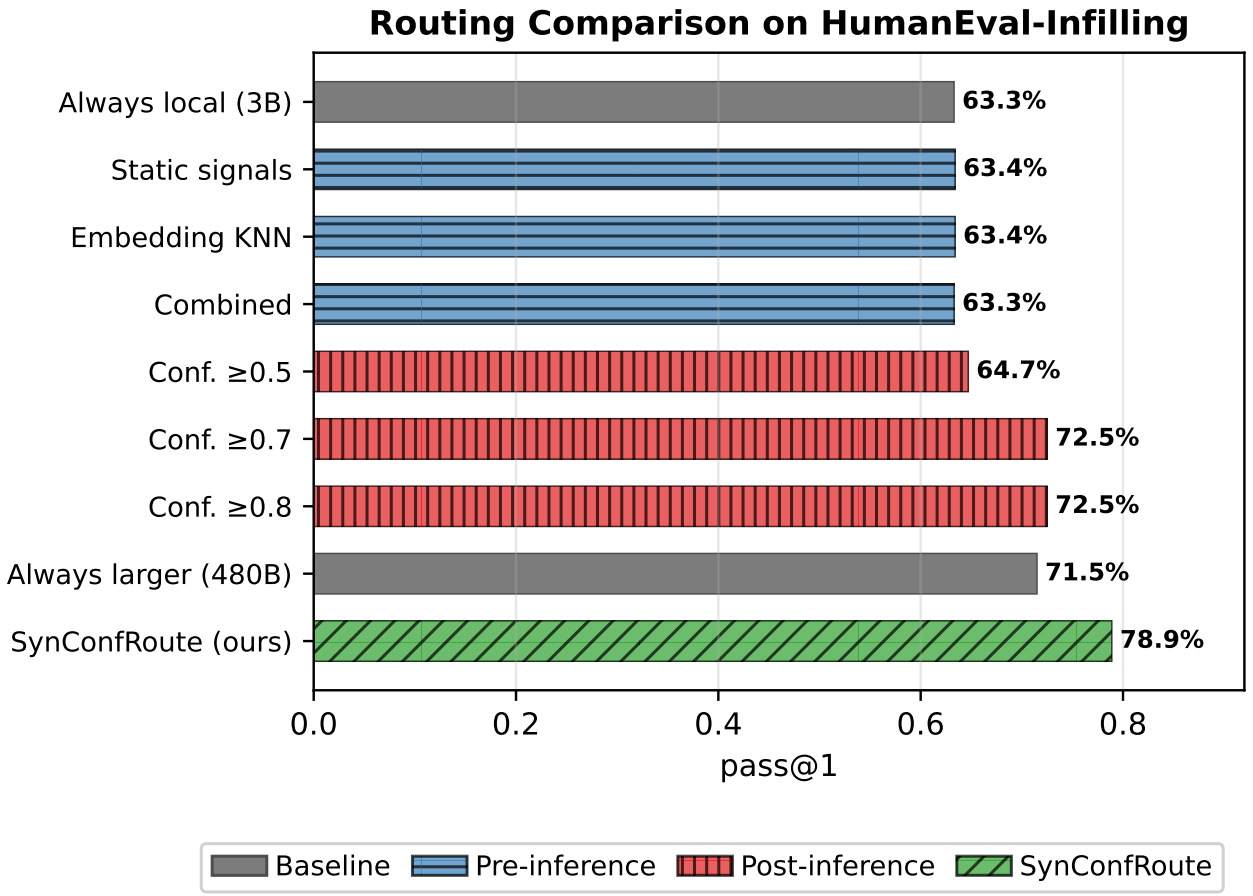

**Figure 3: Routing methods compared by pass@1 and local serving rate (3B local, 480B larger, 833 test tasks). Our *SynConfRoute* (green) achieves 78.9%, outperforming all baselines including the 480B model alone (71.5%).**

72.5% pass@1 with the 3B+480B pair, exceeding both always-local (63.3%) and always-larger (71.5%), while keeping 67.6% of requests on-device. This result may seem surprising: how can routing beat the larger model used alone? The answer is model complementarity. We find that the 3B and 480B models make different errors: 13.7% of solvable tasks are solved only by the 3B model, typically simple return statements where the 480B produces incorrect indentation (e.g., 8-space instead of 4-space indent). Routing preserves these correct local completions instead of unnecessarily replacing them with incorrect 480B outputs. Conversely, 22.0% of tasks are solved only by the 480B, where the 3B model lacks the knowledge to produce the correct completion. Confidence routing captures this complementarity by keeping the 3B model's output when it is confident (and usually correct) and escalating only when confidence is low (indicating likely failure).

**Pre-inference methods cannot exploit complementarity.** Static code signals, embedding KNN, combined routing, and Eagle ELO routing [54] (66.0%) all fail to match confidence routing. These methods decide before seeing the local model's output, so they cannot distinguish tasks the 3B model handles well from those it does not. Pre-inference routing is effective only for very weak local models (0.5B, 1.5B), where blanket escalation to the 32B model lifts pass@1 from near-zero to above 56%.

**Threshold selection is robust.** Over escalation hurts: threshold 0.8 escalates 83.8% of tasks but scores lower than 0.7 (which escalates only 32.4%), because the larger model fails on tasks the 3B solves. The threshold grid search in Algorithm 1 adapts automatically to any model pair. We validate robustness in three ways: *(1) Small calibration sets suffice:* varying calibration size (50, 100, 200, 400 tasks) across three random seeds, the auto-selected threshold yields pass@1 within 0.2% of the 200-task configuration even with only 50 tasks (std <0.4%). *(2) Cross-benchmark transfer:* the threshold calibrated on HumanEval-Infilling generalizes to SAFIM without retuning (escalating 96–98% of tasks where the local model is near-zero). *(3) Cross-family portability:* the algorithm works across model families, not just within Qwen. DeepSeek-Coder-6.7B (29.5% alone) routed to the 480B model reaches 59.1% with the auto-selected threshold, confirming that practitioners can pair any local model with any escalation target.

**Syntax validation catches confident-but-wrong completions.** Among the 166 tasks where the 3B model is confident (≥0.7) but incorrect, 77 (46%) produce syntactically invalid code. Crucially, none of the 397 correct-local completions fail syntax parsing (zero false positives). This result is not obvious a priori: one might expect some correct completions to fail parsing when combined with their surrounding context (e.g., due to indentation mismatches or incomplete surrounding code), which would make syntax checking useless for routing. The zero false positive rate holds across all nine SAFIM splits (three languages, three task types), confirming that this is a robust property, not a benchmark artifact. Adding syntax validation (Algorithm 1) exploits this: at threshold 0.7, *SynConfRoute* achieves **78.9%** pass@1 versus 72.5% for confidence only, a **+6.4%** improvement while keeping 58% local. At threshold 0.6, the improvement is even larger: **78.3%** vs 67.5% (**+10.8%**) with 70% local.

We also tested deeper verification (linting, type checking) and alternative confidence metrics (minimum token probability, average over all tokens); all either introduced false positives or achieved within 0.5% of the current metric. Syntax validation is the sweet spot: zero false positives at <1 ms cost. The remaining 89 confident, syntactically valid, but functionally wrong completions represent the ceiling for static routing; closing this gap requires model-level improvements.

## 4.5 Deployment Strategy Comparison

While Figure 3 compares routing *methods* by pass@1 and local rate, Table 4 evaluates the four deployment *strategies* a practitioner actually chooses between, adding the cost and privacy dimensions that drive real-world adoption decisions.

**Table 4: Deployment strategies on HumanEval-Infilling (833 test tasks). 3B local model, 480B MoE escalation target. Cost is the fraction of requests sent to the shared accelerator (relative to always-larger = 100%); Private is the fraction of completions that never leave the developer's machine.**

| Strategy | pass@1 | Cost | Private |
|---|---|---|---|
| Always local (3B) | 63.3% | 0% | 100% |
| Always larger (480B) | 71.5% | 100% | 0% |
| Conf-only routing | 72.5% | 32% | 68% |
| **SynConfRoute (ours)** | **78.9%** | **42%** | **58%** |
| Oracle (upper bound) | 85.2% | – | – |

*SynConfRoute* achieves the **highest quality** of any strategy (78.9%), including the 480B model used alone (71.5%), while keeping 58% of completions fully private and reducing accelerator cost by 58% relative to always-larger. Confidence-only routing is slightly cheaper (32% vs 42% cost) but at a substantial quality penalty: 72.5% vs 78.9%. In practice, the syntax check adds no latency (<1 ms) and no implementation burden (a single `ast.parse` call), making the +6.4% quality gain effectively free. This is possible because the models make different errors: 13.7% of solvable tasks are solved only by the 3B model, while 22.0% are solved only by the 480B. The 3B-only successes are predominantly simple return statements where the 480B produces incorrect indentation, causing test failures.

**RQ2 Conclusion:** *SynConfRoute* (Algorithm 1) achieves **78.9%** pass@1, a **+6.4%** improvement over confidence only routing (72.5%), with zero false positives and <1 ms cost.

**Implication:** Code completion failures have exploitable structure unavailable in NLP routing. Adding a single syntax check to confidence routing is effectively free and catches 46% of confident errors. RQ3 tests the boundaries of this approach.

## 5 RQ3: Where Does Local-First Routing Work and Where Does It Break Down?

RQ2 established that *SynConfRoute* achieves 78.9% pass@1 on routine single-line completions (HumanEval-Infilling). We now test where this approach breaks down, whether it generalizes across languages, and whether inference-time alternatives can substitute for routing.

### 5.1 Multi-Language Generalization

On SAFIM competitive programming tasks, the 3B model is much weaker than on HumanEval-Infilling. We evaluate *SynConfRoute* across all three SAFIM languages using language-specific syntax checkers: Python's `ast.parse` (<1 ms), Java's `javalang` parser (<1 ms), and `g++ -fsyntax-only` for C++ (~150 ms). Figure 4 shows the results across all nine splits.

*SynConfRoute* improves over confidence only routing on all nine splits across all three languages, with zero false positives. The gains are largest on control tasks (+9–12%) and API tasks (+1–31%), where the 3B model produces syntactically invalid code. Block tasks show smaller gains (+0.4–2%) because failures are predominantly semantic, not structural. In the extreme case (C++ API), all 52 completions from the 3B model are syntactically invalid, so *SynConfRoute* escalates 100% of requests, matching always-larger exactly. This is the correct behavior: when the local model cannot produce valid code, *SynConfRoute* degrades gracefully to always-larger rather than accepting broken completions. This pattern is consistent across Python, Java, and C++: the boundary of *SynConfRoute* is determined by error type (structural vs. semantic), not by programming language.

### 5.2 Why Inference-Time Alternatives Cannot Substitute

An alternative to routing is to improve the local model at inference time. We evaluate two approaches.

**Few-shot retrieval degrades the best model.** Retrieval helps weak models but hurts strong ones: the 1.5B model improves from 12.0% to 20.5% with TF-IDF, but the 3B model drops from 63.3% to 53.3% (−10.0%) and the 480B from 71.5% to 70.8%. Analysis of 178 regressions on the 3B model shows 52% are indentation-copying errors and 28% are content-copying errors where the model mimics the example rather than solving the task.

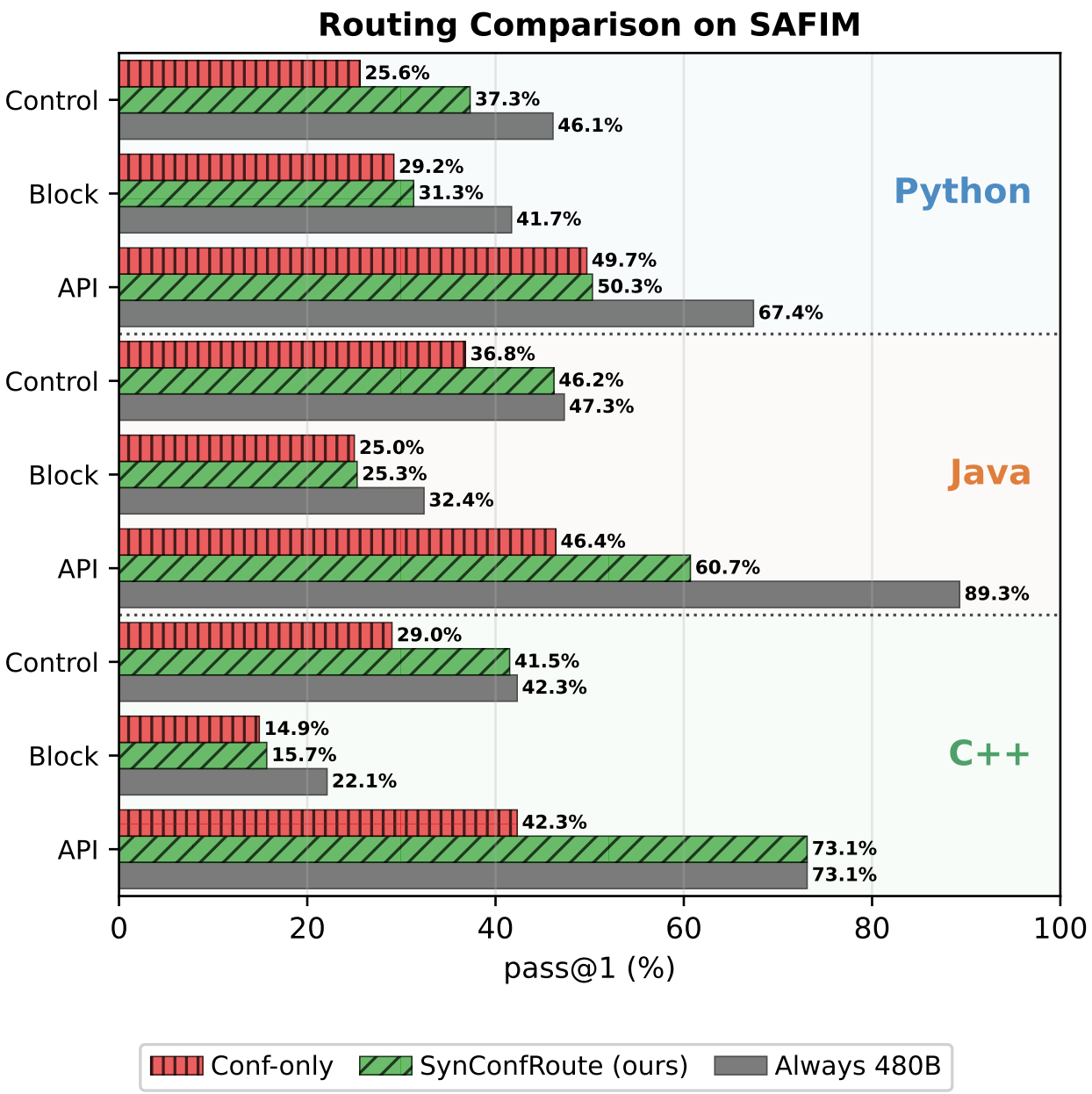

**Figure 4: Routing comparison on SAFIM across Python, Java, and C++ (3B local, 480B larger). *SynConfRoute* (green) improves over confidence only (red) on all nine splits with zero false positives.**

**Multi-sample voting provides negligible gain.** The best configuration (vote@10 at t=0.6) reaches 13.5% pass@1 for the 1.5B model, only +1.6% over greedy (11.9%). The model produces 5.2 unique completions per task on average, but for 866 of 910 failing tasks, none is correct: the model lacks the knowledge to reach the right completion regardless of sampling diversity.

### 5.3 The Remaining Gap

The 54% of confident failures that are syntactically valid but semantically wrong (RQ2) represent the ceiling for static routing. We tested additional signals (linting, type checking, alternative confidence metrics, multi-model agreement) but none improved over *SynConfRoute* without introducing false positives. Closing this gap requires model-level improvements rather than routing enhancements.

**RQ3 Conclusion:** *SynConfRoute* generalizes across Python, Java, and C++, improving over confidence only on all nine SAFIM splits without ever rejecting a correct completion. The boundary is error type (structural vs. semantic), not language.

**Implication:** Inference-time alternatives (few-shot retrieval, multi-sample voting) cannot substitute for routing because the small model's failures are knowledge gaps, not sampling issues. The pipeline is immediately deployable across languages using off-the-shelf parsers.

# 6 Discussion

This section synthesizes findings across the three research questions and discusses their practical implications.

## 6.1 Positioning Against Current Industry Practice

Organizations deploying AI code completion today choose among three strategies, each with a fundamental limitation: (1) *External API* (e.g., GitHub Copilot): high quality but source code leaves the organization, creating privacy and compliance risk; (2) *Self-hosted large model*: private but requires expensive shared accelerator infrastructure for every request; (3) *Local-only small model* (e.g., JetBrains Full Line Code Completion [43]): fully private and low-latency but lower quality on harder tasks. A hybrid approach that combines the strengths of (2) and (3) is possible: routine completions stay local (private, fast, free), while only the tasks the local model cannot handle are escalated to the organization's self-hosted large model (no external API). *SynConfRoute* improves on existing routing methods by adding syntax validation to catch failures that confidence alone misses, achieving higher quality than any single model (+7.4% over the 480B alone), 58% lower infrastructure cost than always-larger, and 58% of completions fully private.

## 6.2 Local-First Deployment Architecture

Our results validate the local-first architecture introduced in RQ2 (Figure 2).

**Tier 1: Local model, 0-shot.** Qwen2.5-Coder-3B at Q4_K_M runs on the developer's workstation accelerator, achieving 64.7% pass@1 at 0.15 s latency with no data leaving the developer's machine. The quantized 3 B model requires only ~2 GB of accelerator memory, fitting comfortably on any workstation accelerator with 8 GB or more.

**Tier 2: SynConfRoute escalation.** When the local model's confidence is low or its completion is syntactically invalid, the request escalates to a larger model. With the 480B MoE model, this achieves 78.9% pass@1 while keeping 58% of completions in Tier 1. The full escalation pipeline completes in 0.32 s (0.15 s local inference + 0.17 s 480B inference) when the small model runs on accelerator.

**SynConfRoute adapts to task difficulty automatically.** The router does not require task-type labels or complexity annotations: confidence and syntax validity together act as an implicit difficulty estimator. On routine completions (single-line assignments, returns, simple expressions), the local model is confident and produces valid syntax, so the request stays local. On harder tasks (complex algorithms, multi-line logic), the local model's confidence drops or its completion fails to parse, triggering escalation. This means developers do not need to decide when to use the small model versus the large model; *SynConfRoute* makes that decision per-request. On our benchmarks, this local-first strategy reduces shared accelerator usage by 58% compared to routing every request to the large model, while achieving higher quality than the large model alone.

Crucially, the entire pipeline uses off-the-shelf models served via standard frameworks with no custom training, making it immediately deployable. Table 5 summarizes the recommended configurations.

**Table 5: Recommended deployment configurations using Qwen2.5-Coder-3B at Q4_K_M (Tier 1). 480B uses *SynConfRoute*; 32B uses confidence only. Both large models served on 8 accelerators; the 480B MoE is faster due to sparse activation. Routed latency = local + escalation.**

| Scenario | Setup | Latency | pass@1 |
|---|---|---|---|
| Private (local) | 3B on wkst. accel. | 0.15 s | 64.7% |
| Quality (32B esc.) | 3B+32B routed | 0.85 s | 70.0% |
| Quality (480B esc.) | 3B+480B routed | 0.32 s | 78.9% |

**Practitioner deployment guide.** (1) Install Ollama and pull a quantized 3B model (~2 GB). (2) Serve the escalation target (32B or 480B) via vLLM on an organization accelerator. (3) Calibrate the confidence threshold on 50+ held-out tasks using grid search. (4) Add a syntax checker for each target language (`ast.parse` for Python, `javalang` for Java, `g++` for C++). (5) Route per-request using Algorithm 1. No training, fine-tuning, or labeled data is required. While our evaluation uses benchmarks rather than a live IDE, it validates every component a deployment requires: model inference latency on workstation hardware (0.15 s), memory footprint (~2 GB), routing decision accuracy (zero false positives), and end-to-end quality (pass@1 via execution). The remaining validation gap is user acceptance behavior, which requires an IDE plugin study and remains future work.

## 6.3 Latency in Practice

Production code completion systems target sub-200 ms latency for synchronous inline suggestions [33, 43]. On a workstation accelerator, the recommended 3B model completes in **0.15 s**, with the 1.5B and 7B models also under 200 ms (0.16 s and 0.19 s). Of the 23 small and mid-range models we evaluated on accelerator, 11 meet the 200 ms target.

## 6.4 Model Selection Over Model Size

Practitioners should prioritize model family over parameter count, and use Q4_K_M quantization (best quality-latency trade-off for the 3B model; models below 3B should use q8_0). This recommendation is validated across 29 models from 12 families: the 3B model outperforms all competitors on both HumanEval-Infilling and SAFIM, including models 2–3× its size (CodeGeeX4-9B, Yi-Coder-9B). As new code models are released, specific model recommendations may change. We verified this by evaluating Qwen3-Coder-Next [39], a newer 80B MoE model with 3B active parameters trained for agentic coding: despite its larger capacity, it achieves only 41.1% on HumanEval-Infilling, well below the FIM-trained Qwen2.5-Coder-3B (64.7%), confirming that training objective (FIM vs. agentic) matters more than model scale for line-level completion. Our evaluation methodology and *SynConfRoute* are model-agnostic: practitioners can re-run the benchmark on any new model, and the threshold calibration adapts automatically to any model pair.

## 6.5 Why Augmentation Cannot Substitute for Routing

As shown in RQ3, inference-time alternatives fail because the small model's failures are *knowledge gaps*, not sampling or context issues. Retrieval causes the model to copy examples rather than solve tasks; voting explores plausible-but-wrong completions. Even cross-model retrieval (using 480B completions as examples for the 3B model) degrades quality: 51.4% with TF-IDF vs 63.3% at 0-shot [49]. Routing succeeds where augmentation fails because it delegates hard tasks to a model that actually has the knowledge.

## 6.6 What Overturns Common Assumptions

Several findings are specific to the FIM code completion domain and do not follow from prior NLP studies. Parameter count is misleading (3B > 7B within the same family). Few-shot retrieval degrades the best small model by 10%. Most importantly, code completion failures have a structure that NLP text lacks: confident-but-wrong completions decompose into structurally broken (46%, catchable by syntax checking) and semantically wrong (54%, not catchable). This decomposition is impossible in NLP routing, where no cheap structural validity check exists. Prior routing work (FrugalGPT [5], RouteLLM [35], Eagle [54], MCCom [28]) treats model outputs as opaque text and routes based on confidence, learned features, or user behavior. Code completion is, to our knowledge, the first domain where structural validity has been exploited as a routing signal, opening a new class of signals beyond token confidence.

# 7 Related Work

Our work builds on five areas: code language models, code completion benchmarks, model routing, few-shot retrieval, and quantization.

**Code Language Models.** Recent CodeLLMs including DeepSeek-Coder [18, 56], StarCoder2 [27], OpenCoder [20], Granite-Code [29], Yi-Coder [1], CodeGemma [9], CodeGeeX4 [55], Magicoder [50], and Qwen2.5-Coder [21] have advanced code generation through FIM training objectives [3], auxiliary objectives [11], and curriculum learning [42]. These models are deployed in industrial IDE systems with strict latency requirements: Meta's CodeCompose [31] fine-tunes CodeLlama for multi-line suggestions, JetBrains ships lightweight language-specific models for low-latency completions [43], and even minor delays disrupt developer workflow [13, 37]. However, hosted systems introduce privacy and data governance concerns [10, 51]. Izadi et al. [22] evaluate code LMs in a real IDE deployment with 600K+ completions, but do not study routing or local deployment. Local frameworks such as Ollama [34] and llama.cpp [15] enable on-device deployment, but quality trade-offs have not been systematically studied.

**Code Completion Benchmarks.** HumanEval-Infilling [3] extends HumanEval [6] to FIM by masking individual lines. SAFIM [17] provides syntax-aware FIM tasks evaluated via ExecEval [23]. Repository level benchmarks (RepoBench [26], CrossCodeEval [12], RepoEval [53]) cover cross-file completion but use surface-similarity metrics. M2rc-Eval [25] provides multilingual repository level evaluation across 18 languages. We use HumanEval-Infilling and SAFIM because both provide execution-based pass@1.

**Model Routing and Cascading.** FrugalGPT [5] introduced LLM cascades for cost reduction. RouteLLM [35] trains router models on preference data. For code, Model Cascading for Code [4] uses self-generated test cases as confidence signals. Most directly relevant is MCCom [28], which cascades a custom-trained 121M local model with a 7B cloud model using token-level confidence and user typing behavior for IDE completion. MCCom evaluates with surface-similarity metrics (exact match, edit similarity) on a workstation accelerator, using a fixed confidence threshold of 0.8. Eagle [54] introduces training-free ELO-based routing; Gatekeeper [40] fine-tunes models for deferral; and a recent survey [30] provides a comprehensive taxonomy. Crucially, no prior routing work uses *code-specific validation signals* such as syntax checking: all existing methods route based on confidence, learned features, or user behavior, ignoring that code correctness can be partially verified before escalation.

Our work differs from MCCom in four ways: (1) we use off-the-shelf CodeLLMs rather than a custom-trained model; (2) we evaluate with execution-based pass@1 rather than surface similarity; (3) we propose *SynConfRoute* (confidence + syntax validation) routing, which improves pass@1 by 6.4% over confidence only by catching syntactically invalid completions that the model is confident about; and (4) we systematically compare seven routing methods on FIM benchmarks across 29 models and three languages.

**Few-Shot Retrieval for Code.** Prior work studies example selection for code synthesis [52], BM25-based repository retrieval [53], and retrieval for local codebases [19]; a recent survey [46] covers retrieval-augmented code generation broadly. Our contribution is showing that the optimal strategy depends on model capability: weak models favor TF-IDF, strong small models are degraded by any example, and only the 32B model benefits from embedding retrieval.

**Quantization for CodeLLMs.** Prior work confirms 4-bit as the safe quantization frontier [16], shows quantization preserves correctness with AWQ [2], and provides token-level analysis [44]. We extend this by evaluating four levels on Qwen2.5-Coder via execution-based FIM, finding q2_K catastrophic and Q4_K_M optimal.

Our study combines these perspectives: 29 models, four quantization levels, two benchmarks, three languages, and seven routing methods, providing practical guidance for local-first IDE completion.

# 8 Threats to Validity

**Internal validity.** We use a unified chat-based prompt rather than native FIM tokens for fair comparison across all 29 models; our best small model via chat (64.7%) outperforms all native FIM configurations. The chat format introduces an indentation repair heuristic that may not generalize to all cases, and limits HumanEval-Infilling to single-line completion (SAFIM includes multi-line tasks, partially mitigating this). Greedy decoding eliminates sampling variance but may underestimate pass@k for k > 1.

**External validity.** Our evaluation uses benchmarks, not a real IDE deployment. Real-world completions involve additional factors (user acceptance, cross-file context) not captured here. We evaluate on Python, Java, and C++; generalization to other languages and repository level scenarios remains future work. Routing is primarily evaluated within the Qwen family to isolate routing from tokenizer mismatches; cross-family portability is validated with one pair (DeepSeek-6.7B→480B). *SynConfRoute* is architecture-agnostic by design, requiring only token log-probabilities and a syntax parser with no model-specific fine-tuning. The 54% of confident failures that are semantically wrong but syntactically valid represent its ceiling; closing this gap requires model-level improvements.

**Construct validity.** Our study focuses on inline code completion (single-line and short multi-line FIM), not other IDE tasks such as code generation or refactoring. These tasks require different model capabilities (e.g., instruction following, long-context reasoning) and may favor different models than those recommended here. In practice, an IDE may use separate models for different tasks; whether *SynConfRoute* generalizes to routing for these other tasks is an open question. Pass@1 measures functional correctness but not readability or style. Benchmark choice affects conclusions: HumanEval-Infilling fails to discriminate the 3B and 32B models (64.7% vs 64.8%), while SAFIM reveals a 9.6% gap on harder tasks.

## 9 Conclusion

This paper studied local-first code completion by evaluating 29 CodeLLMs from 12 families and proposing *SynConfRoute*, a training-free routing method that combines confidence thresholding with syntax validation. Our key finding is that code completion failures have an exploitable structure: 46% of confident-but-wrong completions are syntactically invalid, and this signal achieves zero false positives across three languages. By catching these failures, *SynConfRoute* achieves 78.9% pass@1, outperforming both the local 3B model (63.3%) and the 480B model alone (71.5%), while keeping 58% of requests on-device.

For practitioners, the actionable recommendation is: deploy a well-trained 3B model at Q4_K_M quantization (~2 GB) with *SynConfRoute* escalation to a self-hosted larger model. The entire pipeline uses off-the-shelf models and standard serving frameworks with no custom training.

Future work should validate this architecture in a real IDE deployment with user acceptance studies, extend routing to additional languages, and explore model-level improvements to close the remaining gap to the oracle (85.2%).

## Disclaimer

Any opinions, findings, conclusions, or recommendations expressed in this material are those of the author(s) and do not reflect the views of Huawei. AI tools were used for copy-editing. All experiments, analysis, writing, and results were performed by the authors, who thoroughly reviewed the final content. This complies with IEEE and ACM policies on AI use in publications.

## Data Availability Statement

The replication package for this study is publicly available [47]. It includes all model inference scripts, routing implementations (*SynConfRoute*, confidence only, Eagle ELO, static signals), prediction artifacts for all 29 models across both benchmarks, confidence scores, quantization sensitivity results, and a `README.md` with step-by-step reproduction instructions. All benchmark data (HumanEval-Infilling and SAFIM) is publicly available from their respective repositories.